\documentclass[10pt,letterpaper,twocolumn]{article}

\usepackage{ol2}
\usepackage[draft]{hyperref}
\usepackage{amsmath}
\usepackage{caption}

\begin{document}

\twocolumn[

\title{\textbf{Non-additivity in laser-illuminated many-atom systems}}

\author{Ephraim Shahmoon$^1$, Igor Mazets$^{2,3}$ and Gershon Kurizki$^1$}
\address{
$^1$Department of Chemical Physics, Weizmann Institute of Science, Rehovot, 76100, Israel
\\
$^2$Vienna Center for Quantum Science and Technology, Atominstitut, TU Wien, 1020 Vienna, Austria
\\
$^3$Ioffe Physico-Technical Institute of the Russian Academy of Sciences, 194021 St. Petersburg, Russia
}

\begin{abstract}
We show that atoms subject to laser radiation may form a non-additive many-body system on account of their long-range forces, when the atoms are trapped in the vicinity of a fiber with a Bragg grating.
When the laser frequency is inside the grating's bandgap but very close to its edge, we find that the range and strength of the laser-induced interaction becomes substantially enhanced, due to the large density of states near the edge, while the competing process of scattering to the fiber is inhibited. The dynamics of the atomic positions in this system conforms to a prominent model of statistical physics which exhibits slow relaxation. This suggests the possibility of using laser-illuminated atoms to study the characteristics of non-additive systems.
\end{abstract}


 ]

\emph{Introduction.} Studies of many-body systems with long-range interactions, where the pairwise potential decays at large distances as $1/r^{\alpha}$ with $\alpha$ equal or smaller than the space dimension, have offered in recent years exciting new insights into fundamental concepts of statistical physics \cite{BOOK,MUK}. A direct consequence of long-range interactions is the non-additivity of energy, namely, that the total energy of a system with $N+M$ particles $E(N+M)$, cannot in general be represented by the sum of energies of its subsystems $E(N)+E(M)$ for any system configuration, due to non-negligible interaction energy between the subsystems. This property was shown to lead to interesting thermodynamic and dynamical effects such as inequivalence of statistical ensembles, negative specific heat, breaking of ergodicity and slow relaxation \cite{BOOK,MUK}, most of which have never been verified experimentally. Physical examples for systems with long-range interactions are found in e.g. astrophysics, magnetism, plasmas and free electron lasers  \cite{MUK}, and in the recent realization of one-dimensional gravity-like attraction between atoms, driven by incoherent absorption of laser light \cite{BAR}.

Another interesting realization is that of \emph{controlled} long-range forces between polarizable atoms or molecules coupled by laser-induced dipole-dipole interaction (LIDDI) \cite{MQED,LIDDI}. In free-space, LIDDI may give rise to a $1/r$ gravity-like inter-atomic potential, leading to roton-like excitations in a Bose-Einstein condensate \cite{OD2}. When the atoms are trapped in the vicinity of an optical fiber \cite{KIM} and free to move along its axis as in \cite{RAU}, the fiber-mediated LIDDI can effectively become one-dimensional (1d): it extends to any range and the atoms may self-organize \cite{CHA}.
A limitation on the possible observability of these effects however, is the scattering of laser photons by the atoms, which, owing to its random spontaneous nature, gives rise to a diffusive atomic motion on top of that affected by the LIDDI \cite{CCTn}. The difficulty is that the energy of LIDDI and this scattering rate are typically comparable, as they both scale like $\Gamma \Omega^2/\delta^2$, where $\Gamma$ is the spontaneous emission rate, $\Omega$  the laser Rabi frequency and $\delta$ its off-resonance detuning.

This paper has two linked objectives. The first is to show that the foregoing limitation can be overcome by LIDDI of atoms in the vicinity of a fiber-Bragg grating (FBG), namely, a 1d photonic bandgap crystal \cite{OKA}.
Our main idea is to tune the laser frequency very close to the bandedge but still inside the bandgap, such that the scattering into the fiber modes is drastically reduced and scattering only to the non-guided, free-space modes, remains. Then, since at the bandedge the density of fiber modes diverges, inter-atomic forces due to LIDDI, mediated by virtual photons of the FBG modes, prevail over the remaining scattering effects. This results in atomic-motion dynamics dominated by a quasi-1d fiber-mediated LIDDI. The enhancement of the coherent inter-dipolar \emph{forces} along with suppressed scattering is based on previous suggestions for coherent \emph{excitation exchange}  via the underlying enhanced \emph{resonant dipole-dipole interaction} and suppressed spontaneous emission \cite{RDDI,JOHN,MOL}.
The LIDDI we find [Eq. (\ref{U})] persists over hundreds of laser wavelengths $\lambda_L$ although it is mediated by evanescent waves inside the bandgap.

The second objective is to use this LIDDI to realize the non-additive, infinite-ranged interacting $XY$ (2d) spin model [Eq. (\ref{HMF})], which is known to exhibit slow relaxation, diverging as the system size, towards its ferromagnetic equilibrium state \cite{MUK,YAM}, a prediction which has never
been tested experimentally. We discuss the possibility of observing this relaxation dynamics for atoms with the FBG-mediated LIDDI. This may open the way to further studies of non-additivity in the context of atomic systems.

\begin{figure}
\begin{center}
\includegraphics[scale=0.30]{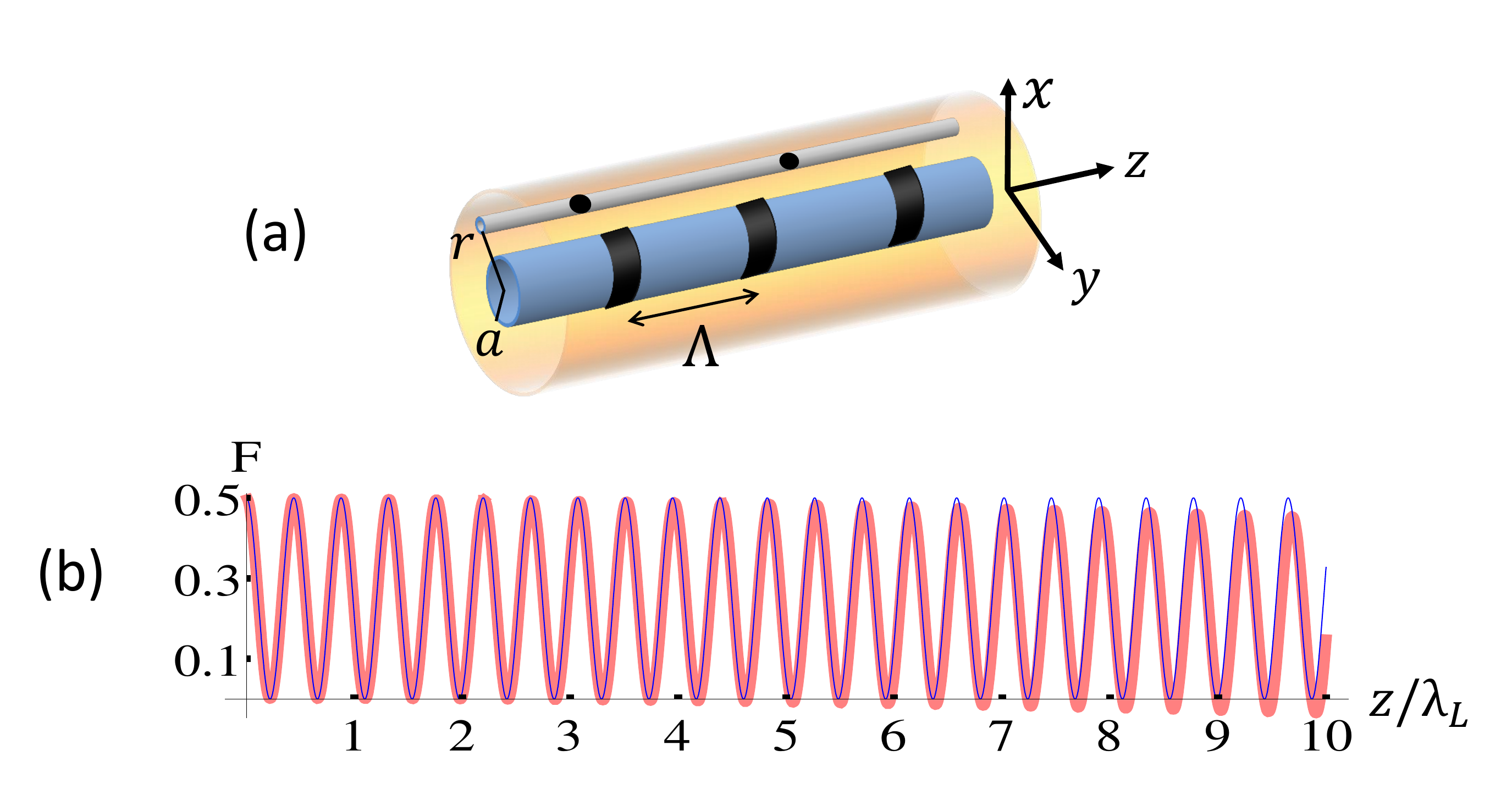}
\caption{\small{
(a) Schematic setup: Atoms trapped in a cylindrical hollow trap along the $z$ axis, and at distance $r$ from the center of the tapered fiber \cite{RAU}. The tapered fiber (central blue cylinder), of radius $a$, refractive index $n_1$ and index grating with period $\Lambda$ and amplitude $\Delta n$, is suspended in vacuum (index $n_0=1$). The guided laser light and fiber modes are coupled to the atoms and induce their mutual interaction (LIDDI).
(b) Spatial dependence of the LIDDI potential: $F(z)$ from Eq. (\ref{U}) (thick pink line) almost coincides with $(1/2)\cos^2(k_L z)$ (thin blue line) up to $z\sim8\lambda_L$. Here $k_L=\bar{n}2\pi/\lambda_L$ (see Cs atoms realization considered in the text).
}}
\end{center}
 \vspace{-0.3em}
\end{figure}

\emph{LIDDI in fiber grating.} Consider the setup depicted in Fig. 1(a), motivated by the experiment of Ref. \cite{RAU}: an ensemble of two-level atoms is trapped in a cylindrical trap outside of a tapered fiber and are free to move along its propagation axis $z$. They are illuminated by off-resonant laser with frequency $\omega_L$ that propagates along $z$ and is assumed to be linearly polarized, i.e. with wavenumber $\mathbf{k}_L=k_L\mathbf{e}_z$ and polarization $\mathbf{e}_L=\mathbf{e}_x$. We can then take the  atomic dipolar transition matrix element as $\mathbf{d}=d \mathbf{e}_x$, i.e. parallel to the laser polarization \cite{LIDDI}.

The dominant long-range LIDDI potential  between the atoms is given in terms of the underlying resonant dipole-dipole interaction, $\Delta_{12}$, as \cite{LIDDI}
\begin{equation}
U=-\frac{|\Omega|^2}{\delta^2} \hbar \Delta_{12} \cos \left[\mathbf{k}_L\cdot(\mathbf{r}_{1}-\mathbf{r}_{2})\right],\,\, \Delta_{12}=\sum_k\frac{g_{k,1}g^{\ast}_{k,2}}{\omega_k-\omega_L},
\label{LIDDI}
\end{equation}
where $\delta$ and $\Omega=E_L\mathbf{e}_L\cdot\mathbf{d}/\hbar$ are the laser detuning (from the atom) and Rabi frequency, respectively, $E_L$ being its electric field amplitude, and $\mathbf{r}_{\nu}$ is the position of atom $\nu=1,2$. $\Delta_{12}$ is mediated by the vacuum of the guided fiber  modes $k$, which determine its spatial dependence via the dipole couplings to atom $\nu$, $g_{k,\nu}=\sqrt{\frac{\omega_k}{2\epsilon_0\hbar}}\mathbf{d}\cdot \mathbf{u}_k(\mathbf{r}_{\nu})$, where $\omega_k$ and $\mathbf{u}_k$ are the modes' frequencies and spatial functions, respectively.
The fiber is made from a core with index $n_1$ and radius $a$ and a clad which is the vacuum with index $n_0=1$.
It has one fundamental transverse mode, the $\mathrm{HE}_{11}$ mode, and various other transverse modes, all having cutoff frequencies, the lowest being $\omega_{co}=2.405(c/a)/\sqrt{n_1^2-n_0^2}$ \cite{OKA}. By assuming the laser wavelength $\omega_L$ and the photonic bandgap (see below) to be much smaller than $\omega_{co}$, we can take only the fundamental $\mathrm{HE}_{11}$ mode into account.
Its mode function has the form $\mathbf{u}_{\beta}=[\mathbf{E}_{\beta}(r,\phi)/\sqrt{A_{\beta}}]e^{i\beta z}/\sqrt{L}$,
where $\beta$ is the wavenumber and $L$ the quantization length on the $z$ axis. The transverse profile $\mathbf{E}_{\beta}(r,\phi)$ for $r>a$, $r,\phi$ being the usual polar coordinates, is given by a linear combination of modified Bessel functions $K_0(q r/a)$, $K_1(q r/a)$ and $K_2(q r/a)$ where we find $q$ and the normalization $A_{\beta}$ for the realization considered below, by first numerically solving for the dispersion relation $\beta(\omega)$ \cite{OKA}. Using these fiber modes we obtain $\Delta_{12}$ as \cite{RDDI}
\begin{equation}
\Delta_{12}=\frac{d^2}{2 \pi \hbar \epsilon_0}\mathrm{P} \int_0^{\infty}d\omega \frac{\partial \beta }{\partial\omega}
\frac{ \cos(\beta z)}{\omega - \omega_L}
\frac{\omega}{A(\omega)}E_x(\omega,\mathbf{r}^{\bot}_{1})E^{\ast}_x(\omega,\mathbf{r}^{\bot}_{2}),
\label{Df}
\end{equation}
with $E_x(\omega)=\mathbf{e}_x\cdot\mathbf{E}(\omega)$, $z=z_1-z_2$ and $\mathbf{r}^{\bot}_{\nu}=(r_{\nu},\phi_{\nu})$ the transverse position of atom $\nu=1,2$. The dispersion relation $\beta(\omega)$ and density of states (DOS) $\partial \beta / \partial\omega$ are used here to express quantities as a function of $\omega$ rather than $\beta$.

Assuming periodic perturbations $\Delta n$ of the refractive index about $n_1$ with period of length $\Lambda$ [Fig. 1(a)], we obtain the FBG dispersion and DOS \cite{RDDI}
\begin{eqnarray}
&&\beta \approx k_B +k_B\sqrt{\bar{n} \Delta n}\times\left\{\begin{array}{c}
                                                                               \sqrt{\omega/\omega_u-1} \quad;\quad\omega>\omega_u\\
                                                                               -\sqrt{1-\omega/\omega_d} \quad;\quad\omega<\omega_d
                                                                             \end{array}\right.,
\nonumber\\
&&\frac{\partial\beta}{\partial \omega}\approx \frac{\sqrt{\bar{n}\Delta n}}{2 c}\times\left\{\begin{array}{c}
                                                                               1/\sqrt{\omega/\omega_u-1} \quad;\quad\omega>\omega_u\\
                                                                               1/\sqrt{1-\omega/\omega_d} \quad;\quad\omega<\omega_d
                                                                             \end{array}\right.,
\label{DOS}
\end{eqnarray}
where $\omega_{u/d}=(c/\bar{n})k_B[1\pm(1/2)(\Delta n/\bar{n})]$ are the upper/lower bandedges, $k_B=\pi/\Lambda$ is the Bragg frequency and $\bar{n}=c k_L /\omega_L$ is the effective refractive index, determined by the dispersion relation without the grating $\beta(\omega)\approx \omega \bar{n}/c$ around $\omega_L$ . The divergence of the DOS near the bandedges gives the main contribution to the integral in (\ref{Df}), and hence allows to use this expression for the DOS in the entire integration range, even though it is valid only for $(\beta/k_B-1)^2(2\bar{n}/\Delta n)^2\ll 1$.

 In order to prevent scattering we take $\omega_L$ to be inside the gap where the DOS vanishes, thus inhibiting photon emission into the fiber modes. However, emission, and hence scattering, to the non-guided, free-space-like modes, still exists and will give rise to diffusive motion of the atoms (see below). Then, in order to make LIDDI the dominant effect (over scattering and diffusion), we consider $\omega_L$ very close to one of the bandedges (but still inside the gap), where the DOS becomes huge, e.g. just below $\omega_u$.
This allows to take  $E_x(\omega,\mathbf{r}^{\bot}_{1}) E^{\ast}_x(\omega,\mathbf{r}^{\bot}_{2}) \omega/A(\omega)$
 out of the integral with $\omega=\omega_u\approx \omega_L$ (an approximation that was verified numerically) and perform the contour integration. Inserting the result into Eq. (\ref{LIDDI}) we finally obtain
\begin{eqnarray}
&&U(z)=- \hbar 2\eta R_{fs}  F(z), \quad R_{fs}=\frac{|\Omega|^2}{2\delta^2}\Gamma_{fs}
\nonumber \\
&&\eta = \frac{\sqrt{\bar{n} \Delta n }}{2\sqrt{1-\omega_L/\omega_u}}  3\pi \left(\frac{c}{\omega_L}\right)^2\frac{E_x(\omega_L,\mathbf{r}^{\bot}_{1})E^{\ast}_x(\omega_L,\mathbf{r}^{\bot}_{2})}{A(\omega_L)},
\nonumber \\
&&F(z)=\cos(k_L z) \left\{\cos(k_B z)\frac{1}{2}e^{-z/l}-
\right. \nonumber\\
&& \left. \sin(k_B z)\frac{1}{2\pi}\left[e^{-z/l}\mathrm{Ei}(z/l)-e^{z/l}\mathrm{Ei}(-z/l)\right]\right\},
\nonumber \\
&&l=\frac{1}{\sqrt{1-\omega_L/\omega_u}}\frac{1}{k_B\sqrt{\bar{n}\Delta n}}.
\label{U}
\end{eqnarray}
Here $R_{fs}$ and  $\Gamma_{fs}=d^2 \omega_L^3/(3\pi\epsilon_0\hbar c^3)$ are the scattering rate and its underlying spontaneous emission rate to free space modes at frequency $\omega_L$ \cite{LIDDI}. Hence, the parameter $\eta$, typically much smaller than 1 for waveguides, measures whether the fiber-mediated LIDDI can become stronger than the scattering to free-space, since $U\propto \eta R_{fs}$.
Let us illustrate the effect of the FBG on $\eta$ for a realization similar to that of Ref. \cite{RAU}, where cesium (Cs) atoms are trapped at a distance $r$ from a fiber with $a=250$nm. Without the grating, for $r=2a,1.5a,1.1a$ we obtain $\eta\sim 0.043, 0.149,0.45$, respectively. With the FBG (see details below), these values considerably improve to $\eta\sim 2.49,8.61,26.1$, respectively, thanks to the $1/\sqrt{\omega_u-\omega_L}$ divergence, thus providing LIDDI stronger than the scattering.

The space-dependence of the above LIDDI potential is encoded in the function $F(z)$, plotted in Fig. 1(b). The second term of this function [in square brackets, see Eq. (\ref{U})] is much smaller than its first term, hence $F(z)$ appears as a beat between two cosine functions exponentially decaying on length scale $l$. Since $l\propto 1/\sqrt{\omega_u-\omega_L}$, it becomes very large at the bandedge, allowing for LIDDI at very long distances. Hence, for system length much smaller than $l$, $F(z)$ is almost purely sinusoidal, and for $z\ll 2\pi/|k_L-k_B|$ it reduces to $F(z)\approx (1/2)\cos^2(k_L z)$. For the Cs atoms example considered below $l\approx 403\lambda_L$ and $2\pi/|k_L-k_B|\approx 101\lambda_L$, such that indeed  $F(z)\approx (1/2)\cos^2(k_L z)$ up to $z\sim 8\lambda_L$, as can be seen in Fig. 1(b).

We conclude that the obtained LIDDI is both strongly enhanced (by a factor $\eta$) and extended in range (with typical scale $l$), whereas scattering to the fiber modes is inhibited inside the bandgap and close to the bandedge. This comes about since $\eta$ and $l$ are both large near the bandedge due to the factor $1/\sqrt{\omega_L-\omega_u}$. The above treatment, based on Ref. \cite{LIDDI}, is perturbative, hence the question of its validity near the bandedge, where the coupling becomes strong, arises. Yet, using a non-perturbative theory as in \cite{RDDI}, we are able to verify that our perturbative analysis is a good approximation in the example considered below.

\emph{Nonadditivity and many-body relaxation dynamics.}
The above long-range LIDDI may open the way to the study of non-additivity in atomic systems. We shall illustrate this idea by the example of slow relaxation dynamics.
The system in Fig. 1(a) \cite{RAU} effectively allows for dynamics only along the $z$-axis since the confinement in the transverse directions is $<60$nm (trap oscillation frequency $\approx140$ kHz \cite{RAU}), much smaller than the scale for spatial variation of the field, $1/q\sim a\sim 250$nm, hence $|\mathbf{r}^{\bot}_{1,2}|\approx r$ is effectively constant. For a system smaller than $l$ and $2\pi/|k_L-k_B|$, $U\propto -\cos^2(k_L z)=(1/2)[-1-\cos(2 k_L z)]$, such that the resulting Hamiltonian for the 1d motion of $N$ atoms of mass $m$ becomes
\begin{equation}
H=\sum_{i=1}^N\frac{p_i^2}{2m} +   \frac{1}{2}\frac{J}{N}\sum_{i,j=1}^N[-1-\cos(\theta_i-\theta_j)],
\quad
\theta_i\equiv 4\pi \bar{n}\frac{z_i}{\lambda_L},
\label{HMF}
\end{equation}
with $J/N\equiv \eta \hbar R_{fs}$ and where $z_i,p_i$ are the 1d coordinate and momentum, respectively, of atom $i$. This is the so-called Hamiltonian mean-field model, describing the infinite-range (non-decaying) coupling between 2d spins with angles $\theta_i$, which is extensively studied in statistical physics of non-additive systems \cite{MUK}. At equilibrium, in both microcanonical and canonical ensembles, this model exhibits a second-order phase transition at temperature $T_c=0.5J$, between a magnetic state, where all spins are aligned, and a non-ordered state \cite{YAM}. In the considered atomic system, the ordered state means that the atoms are placed at lattice sites, $\lambda_L/(2\bar{n})$ apart (see also \cite{CHA}), where the number of atoms at each site is expected to be determined by short-range interactions.
The inequivalence between the microcanonical and canonical ensembles, typical of non-additive systems, is here revealed in dynamics of relaxation towards equilibrium. In the canonical ensemble, where the system is coupled to a bath with coupling rate $\gamma$, the relaxation time $\tau_c$ is typically $\tau_c\sim \gamma^{-1}$. However, in the microcanonical ensemble, where $\gamma\equiv 0$, the long-range interaction gives rise to relaxation times to the ordered equilibrium state, scaling with the system size as $\tau_{\mu c}\sim N^{1.7}$. This is typically revealed for initial configurations where the momentum is distributed uniformly, due to the existence of quasistationary states of the Hamiltonian dynamics, where the system spends a long time prior to relaxation \cite{YAM,ORL}.

Let us now consider the dynamics of the atomic system and then relate it to that described above for the Hamiltonian mean field model. First, we have to take the scattering of laser photons to free-space modes into account, by introducing the proper friction $\gamma$ and the corresponding Langevin force $f_i$, setting $\langle f_i(t) f_j(t') \rangle=2D\delta_{ij}\delta(t-t')$ with $D$ the diffusion constant \cite{CCTn},
\begin{equation}
\gamma\approx-\frac{\hbar k_L^2}{m}\frac{R_{fs}}{\delta}
,\quad D\approx\hbar^2 k_L^2 \frac{R_{fs}}{2}.
\label{dif}
\end{equation}
Defining the time scale $\tau=\sqrt{m/J}\lambda_L/(4\pi)$ and the dimensionless damping $\tilde{\gamma}=\tau \gamma$, temperature $T=D/(m\gamma J)$ and Langevin force $\xi_i=f_i \lambda_L/(4\pi J\sqrt{2\tilde{\gamma}T})$, the equation of motion for $\theta_i$, affected by both the Hamiltonian (\ref{HMF}) and the scattering, then reads,
\begin{equation}
\theta''_i=-\frac{1}{N}\sum_{j=1}^N\sin(\theta_i-\theta_j)-\tilde{\gamma}\theta'_i+\sqrt{2\tilde{\gamma}T}\xi_i,
\label{EOM}
\end{equation}
with $\theta'_i=\tau \dot{\theta}_i$.
The above equation describes two competing relaxation processes: The first term on the right-hand side is that of inter-atomic "collisions" originating from the LIDDI pairwise potential, which ultimately lead, at time $\tau_{\mu c}\sim N^{1.7}$, to a microcanonical relaxation to equilibrium determined by the initial system state. The second and third terms describe however the coupling of the many-atom system to an effective bath due to scattering, which relaxes the system at a canonical-ensemble rate $\tau_c\sim \tilde{\gamma}^{-1}$ to thermal equilibrium determined by the bath temperature $T$.
Clearly, in order to observe the interesting scaling of the slow microcanonical relaxation time with $N$ due to non-additivity, this relaxation has to be faster than that due to the bath, i.e. $\tau_{\mu c}\ll \tau_c $, where we take $\tau_{\mu c}\approx\tau (1/9)N^{1.7}$ \cite{YAM}.
Motivated by the experiment in Ref. \cite{RAU}, we consider the $\Delta m=0$ transition in the D1 line of Cs atoms. The laser parameters are $\delta=-2\pi\times0.65$ GHz and $I=c\epsilon_0E_L^2=10^4$ Wm$^{-2}$, corresponding to $\lambda_L=894.59469$ nm and $\Omega/\delta\approx0.08\ll 1$. For the fiber we take $a=250$ nm, $n_1=1.452$, $n_0=1$, $\Delta n=0.02$ and $\Lambda=396$ nm, resulting in $\bar{n}\approx  1.14$ and $\lambda_u=2\pi c/\omega_u=894.58990$ nm. Fig. 2 presents the comparison between $\tau_{\mu c}$ and $\tau_{ c}$ as a function of $N$ for $r=2a,1.5a,1.1a$. It is seen that for $N$ ranging over several orders of magnitude, the microcanonical slow relaxation is still much faster than the canonical one $\tau_c$, and hence has a chance to be probed by detecting the dynamics of the atomic spatial configuration until it reaches the ordered equilibrium state.

\begin{figure}
\begin{center}
\includegraphics[scale=0.38]{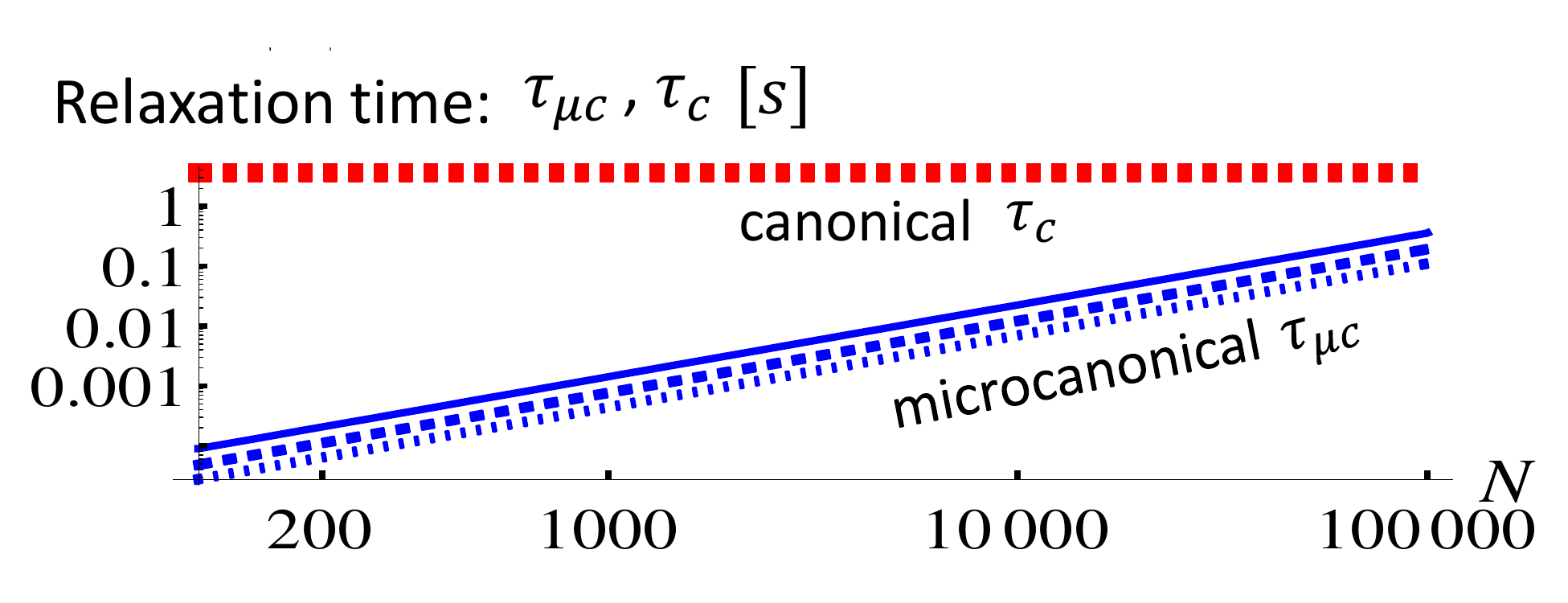}
\caption{\small{
Relaxation time in the microcanonical ensemble $\tau_{\mu c}$ as a function of $N$ for $r=2a,1.5a,1.1a$ (blue solid line, blue dashed line, blue dotted line, respectively), compared to that in the canonical ensemble $\tau_{ c}=\gamma^{-1}$ (red horizontal dashed line) in seconds. The physical parameters used for this plot are in the text. Since $\tau_{\mu c}\ll \tau_{ c}$, $\gamma \neq 0$ does not limit the observation of the scaling of $\tau_{\mu c}$ with $N$.
}}
\end{center}
 \vspace{-0.3em}
\end{figure}

Considering the effect of the Langevin force at finite temperatures $T$ on the relaxation dynamics, it was found by numerical simulations of (\ref{EOM}) that as long as $\tau_{\mu c}\ll \tau_{ c}$, the relaxation is indeed dominated by the microcanonical dynamics \cite{ORL}. Hence, Fig. 2 in fact suggests that the effect of non-additivity, namely, the scaling $\tau_{\mu c}\propto N^{1.7}$, can be revealed by measuring the relaxation time of the considered atomic system for a range of $N$-values.

\emph{Conclusions.}
In this work, we have first derived the pairwise interaction potential (LIDDI) between atoms that are trapped near a fiber grating, and have shown it to be both long-range and stronger than the scattering rate, for a laser frequency inside the bandgap and very close to its edge.
For systems extending over a few laser wavelengths $\lambda_L$, this potential is effectively infinite-ranged, which may open a new route in the study of non-additive effects in atomic and optical physics. This was illustrated by the estimation that the slow relaxation of the Hamiltonian mean-field model, that has never been observed, can in principle be probed in realistic atom-fiber systems. The technical issue of the preparation of the required uniform initial atomic momentum distribution \cite{YAM,ORL} may be resolved by using Raman-transition-based velocity selection techniques \cite{CHU} in order to, e.g. cut off the high-momenta edges of a thermal distribution.
The system relaxation process may be detected by accumulating the
statistics of the correlation properties of an atomic cloud released from a trap  after different holding times using the atom-cloud focusing method \cite{SHL} or by reflection spectra \cite{CHA}, whereas the ability to create a tapered-fiber FBG was recently demonstrated \cite{HAK}.

We stress the generality of the approach utilized here for designing systems with long-range interactions, manifest in Eq. (\ref{LIDDI}) for the LIDDI potential by $\Delta_{12}$, which depends on the spatial structure of the photon modes, and hence on geometry. Other geometries may give rise to hitherto unexplored nonadditive systems.
\\\\
We acknowledge the support of ISF, BSF and FWF (Project No. P25329-N27),
and fruitful discussions with A. Rauschenbeutel, D. Mukamel and O. Hirschberg.

\newpage
\textbf{\emph{Full references}}
\\\\

\end{document}